\documentclass{aa}
\usepackage[utf8]{inputenc}
\usepackage{graphicx}        
\usepackage{natbib}         
\usepackage{amssymb}        
\usepackage{amsmath}
\usepackage{color} 
\usepackage{MR_AA}
\graphicspath{{./}}

\newcommand{\rsun}{\mbox{\rm R$_\sun$}}

\newcommand{\Req}{R_{\mathrm{eq}}}
\newcommand{\Rpol}{R_{\mathrm{pole}}}
\newcommand{\Teq}{T_{\mathrm{eq}}}
\newcommand{\Tpol}{T_{\mathrm{pole}}}

\newcommand{\veq}{v_{\mathrm{eq}}}

\title{Spectroscopic detection of Altair's non-radial pulsations\thanks{Based on observations obtained at the T\'elescope Bernard Lyot
(TBL) at Observatoire du Pic du Midi, CNRS/INSU and Universit\'e de
Toulouse, France.}}

\author{Michel Rieutord\inst{1}
\and Pascal Petit\inst{1}
\and Daniel
Reese\inst{2}
\and Torsten Böhm\inst{1}
\and Arturo L\'opez Ariste\inst{1}
\and\\
Giovanni M. Mirouh\inst{3} \and Armando Domiciano de Souza\inst{4}}

\institute{
IRAP, Universit\'e de Toulouse, CNRS, UPS, CNES,
14, avenue \'{E}douard Belin, F-31400 Toulouse, France
\and 
LESIA, Observatoire de Paris, Université PSL, CNRS, Sorbonne Université, Univ. Paris Diderot, 				 Sorbonne Paris Cité, 5 place Jules Janssen, 92195 Meudon, France
\and
Departamento de F\'isica Te\'orica y del Cosmos, Universidad de Granada, Campus de Fuentenueva s/n, E-18071 Granada, Spain
\and
Université Côte d'Azur, Observatoire de la Côte d'Azur, CNRS, Laboratoire Lagrange, Bd de l'Observatoire, CS 34229, 06304 Nice cedex 4, France
\\
\email{[Michel.Rieutord,Pascal.Petit,tboehm,Arturo.Lopezariste]@irap.omp.eu,daniel.reese@obspm.fr, \\
gmm@ugr.es,armando.domiciano@oca.eu}
}

\date{\today}

\begin{document}

\abstract
{Rapid rotation is a common feature of early-type stars but which
remains a challenge for the models. The understanding of its effect on
stellar evolution is however imperative to interpret the observed
properties of numerous stars.}{We wish to bring more
observational constraints on the properties of fast rotating stars,
especially on their oscillation modes.}{We focus on the nearby star
Altair which is known as a very rapidly rotating star with an equatorial
velocity estimated recently at 313 km/s. We observed
this star with the high-resolution spectropolarimeter
Neo-Narval during six nights, with one night of interruption, in September
2020.}{We detect significant line profile variations on the mean line
profile of the spectra. Their time-frequency analysis shows that these
variations are induced by gravito-inertial waves propagating at Altair's
surface with azimuthal wavenumbers of order $m=10-15$. With a preliminary
computation of the eigenspectrum using the most recent concordance model
of Altair we can give a first modelling of the observed waves.}{
Altair was known as the brightest $\delta$ Scuti star.  We now see that
it is the brightest hybrid oscillating star with excited gravito-inertial
waves and acoustic waves.  Clearly, more observations and more advanced
models are needed to explain the observations in greater details.}


\maketitle

\section{Introduction}

At a distance of 5.13~pc, Altair ($\alpha$ Aql) is the closest
early-type (A7V) fast rotating star to the Sun.  As such,
Altair has been intensively observed in interferometry
\cite[][]{vanbelle+01,domiciano+05,petersonetal06a,monnier+07,bouchaud+20,spalding+22}.
An important result of interferometric observations is the determination
of the inclination of Altair's rotation axis on the line-of-sight. The
most recent value is $i=50.7\pm1.2^\circ$ \citep{bouchaud+20}. Since
spectroscopy indicates that  $v\sin i\simeq 227\pm11$ \citep{reiners+04b},
$231$~km/s \citep{takeda2020}, $242$~km/s \citep{bouchaud+20}, it turns
out that Altair's equatorial velocity is likely over 300~km/s. This makes
Altair rotating at 74\% of its critical angular velocity. As a consequence
it is strongly flattened by the centrifugal force, making its equatorial
radius $22$~\% larger than the polar one. Hence, it is no surprise that such
a star cannot be modelled with spherically symmetric models as is shown by
its age undetermination. Indeed, using 1D-models \cite{suarez+05} indicate
a range of 225-775~Myrs, while \cite{domiciano+05} mention another range
in between 1.2 and 1.4 Gyrs. Thus, Altair turns out to be the star to be
used for testing 2D models like ESTER models \citep{ELR13,RELP16}. This
test was the main objective of \cite{bouchaud+20} who devised the first
concordance model of Altair. \cite{bouchaud+20} indeed managed to match
the interferometric, spectroscopic and asteroseismic observations with
a single 2D-model. This model indicates that Altair is 100~Myrs old
thus barely off the ZAMS as suspected by \cite{petersonetal06a}. In
this 2D-modelling, \cite{bouchaud+20} showed that asteroseismology
was key to lift some degeneracy on the mass of Altair. Altair is
indeed a $\delta$ Scuti star whose oscillations were first detected
by \cite{buzasietal05}. Recently, \cite{ledizes+21} confirmed these
$\delta$ Scuti oscillations and increased slightly the number of
detected frequencies, thanks to the analysis of MOST (Microvariability
and Oscillations of STars) satellite data \citep{walker+03}. \cite{ledizes+21}
also showed the variability of the modes amplitudes, and
the probable coupling of the modes with thin convective layers not much
below Altair's surface.

In the present work we describe the first detection of non-radial
pulsations of Altair with high resolution spectroscopy using the 2-meter
Bernard Lyot telescope at the Pic-du-Midi. In the past similar detections have
been made on very few rotating stars: e.g. $\gamma$ Bootis with
$v\sin i \simeq 127$~km/s \citep{ventura+07}, or on spectroscopic binaries
like RS Chamaeleontis \citep{bohm+09}. However, Altair is the star with
the largest $v\sin i$ where such oscillations are detected. Compared to
photometric observations, spectroscopic ones give an indication of the
mode azimuthal wavenumber which helps with mode identification and further
constraining the fundamental parameters.

The paper is organised as follows: We first give a brief description  of
the data we use and their reduction (Sect. 2). This is followed by their
analysis (Sect. 3) and a short discussion of the possible identification
of modes (Sect. 4). We then revert to 2D models and discuss the comparison
between data and model predictions (Sect. 5). Conclusions follow.

\section{Observations and data reduction}

Altair was observed during six nights in between 2 September and 8
September 2020 at the Pic-du-Midi with the spectropolarimeter Neo-Narval
\cite[see][for a presentation of the instrument]{lopezariste+22} at the Cassegrain focus of the Bernard Lyot Telescope. 643
spectra were obtained sampling the light of Altair every 3~min or so
(see Tab.~\ref{obs_date} for details). The duty cycle of these
observations is however quite low at 0.21.

From the circular and linear spectropolarimetric data collected during
this run, only the intensity signal was used. The data reduction was
performed through the automated pipeline of the instrument, providing
us with spectra covering the whole optical domain (380 to 1,050 nm)
at a resolving power of around 65,000. The peak signal-to-noise ratio,
defined per unit of spectral resolution,
is typically close to 1,400. This latter value is reached at wavelengths
close to 705~nm.

\begin{figure}
\includegraphics[width=0.95\linewidth]{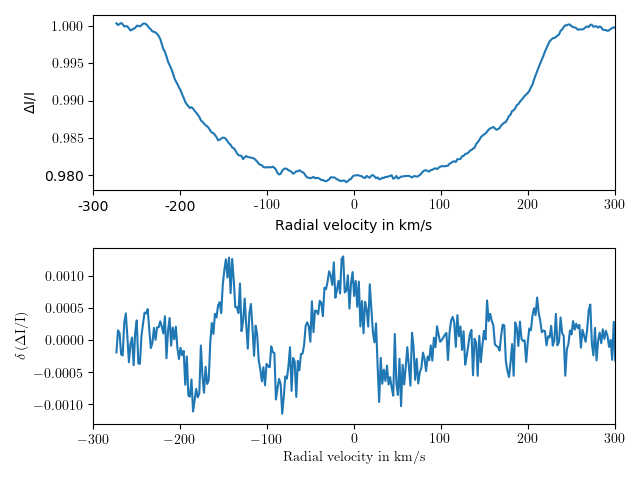}
\caption{Upper panel: example of an LSD profile of Altair's
spectrum. Bottom panel: the same LSD profile, after subtraction of the
average of all available profiles of the same night.}

\label{mean_prof}
\end{figure}

All observations were processed with the Least-Square Deconvolution
method \citep{donati+97,kochukhov+10} to extract a mean pseudo-line
profile from every spectra. A list of around 6,300 photospheric
lines deeper than 1\% of the continuum was extracted from the VALD
data base \citep{kupka+99}, assuming a surface temperature equal
to 7,500~K. According to \cite{monnier+07}, this temperature is
representative of intermediate latitudes of this oblate star. The
intensity signatures reported hereafter are qualitatively unchanged when
adopting other line masks with temperatures chosen to match lower or
higher latitudes. The LSD profile example shown in Fig.~\ref{mean_prof}
(top) is dominated by rotational broadening. Once an averaged profile (of all
observations collected during the same night) is subtracted, bumps and
dips become visible within line profiles (Fig.~\ref{mean_prof} bottom).

\begin{table}[t]
    \centering
\begin{tabular}{|c|c|c|c|}
Night & JD$_s$-JD$_0$  &  JD$_e$-JD$_0$ & $\moy{\delta t}$ \\
      &   day    &   day      &     seconds \\
2-3 Sept. 2020 & 1.40711 & 1.59317 & 206  \\
3-4 Sept. 2020 & 2.31025 & 2.57120 & 211  \\
4-5 Sept. 2020 & 3.31218 & 3.56229 & 165  \\
5-6 Sept. 2020 & 4.32412 & 4.57411 & 162  \\
7-8 Sept. 2020 & 6.46547 & 6.55644 & 160  \\
8-9 Sept. 2020 & 7.30241 & 7.56421 & 162  \\
\end{tabular}
\caption{Summary of the observations dates and the average period of time
sampling $\moy{\delta t}$. JD$_s$ and JD$_e$ are the starting and ending
Julian dates of the observations.
The reference Julian date is JD$_0$=2459094.}
    \label{obs_date}
\end{table}

\begin{figure}
\includegraphics[width=0.95\linewidth]{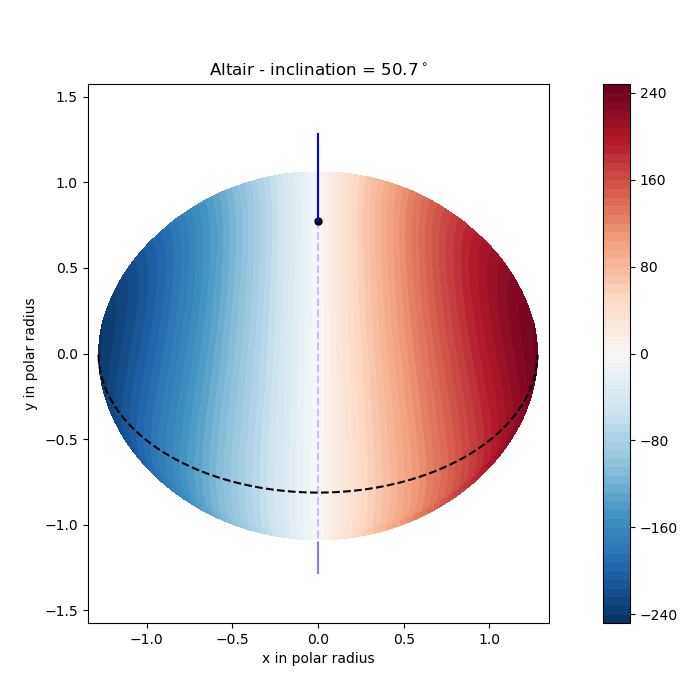}
\caption{The rotational velocity field of Altair projected along the line
of sight according to the model of \cite{bouchaud+20}. Velocities are
in km/s. The equator and North pole are marked out.}
\label{vproj}
\end{figure}

\section{Data analysis}

Fig.~\ref{vproj} illustrates the way we see the Doppler shift of Altair
according to the model of \cite{bouchaud+20}.  This model also gives
an equatorial rotation frequency of 3.08 c/d \citep{ledizes+21}, which
we shall adopt when moving from the observer's frame to the co-rotating
frame. Any feature  moving in the line profile may or may not move in
longitude in the co-rotating frame.

\begin{figure*}
    \centering
    \includegraphics[width=0.30\linewidth]{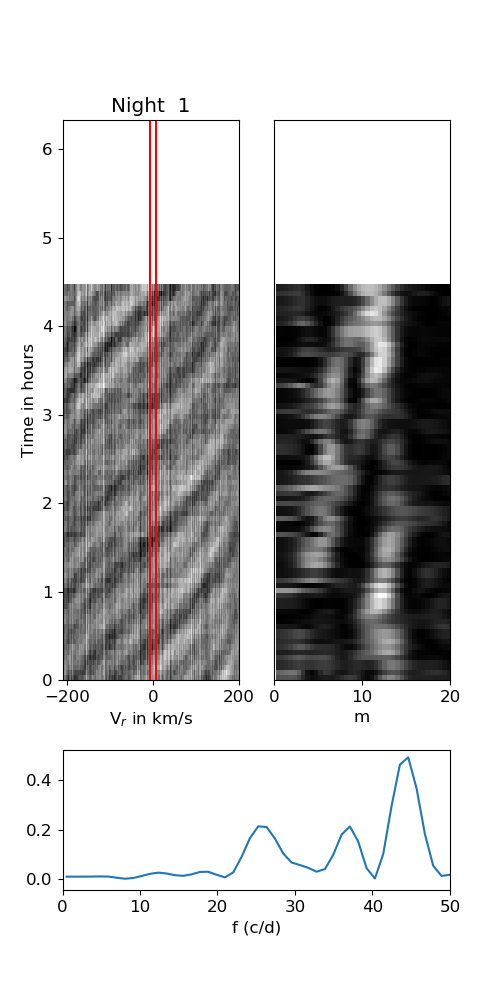}
    \includegraphics[width=0.30\linewidth]{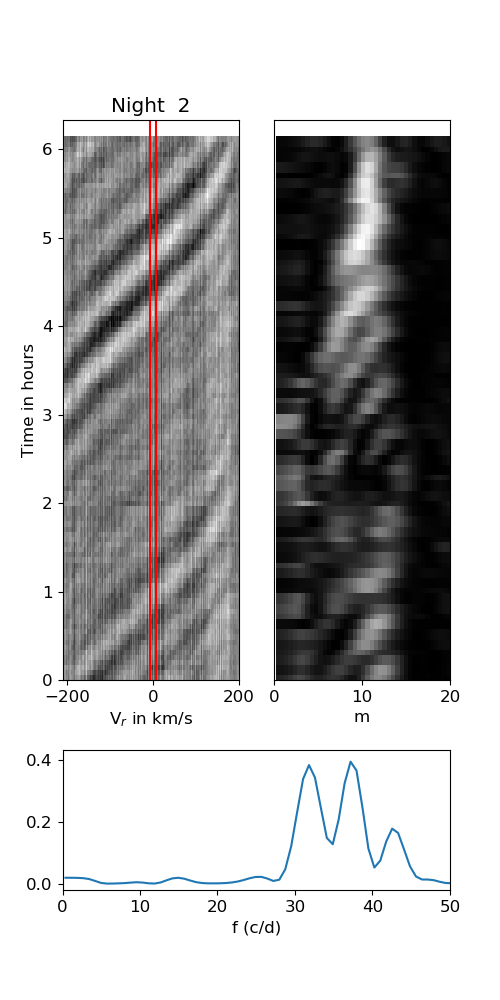}
    \includegraphics[width=0.30\linewidth]{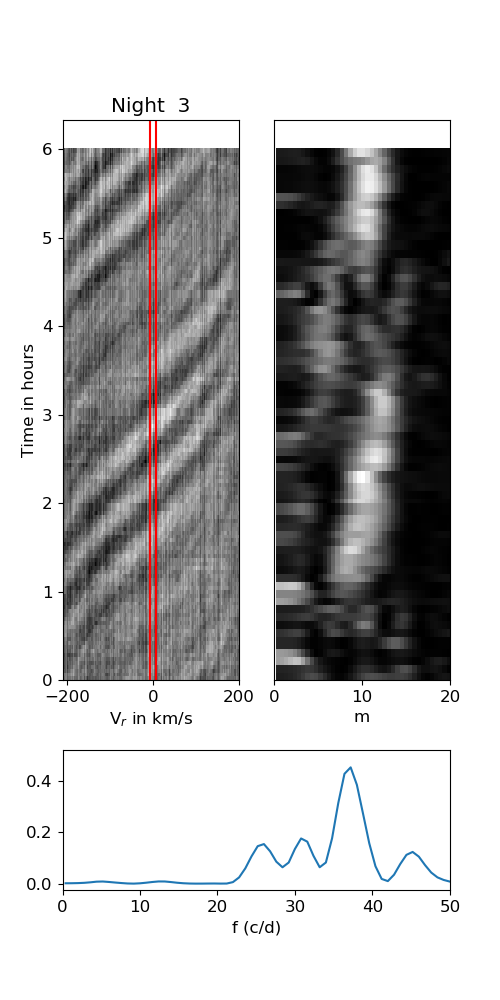}
    \includegraphics[width=0.30\linewidth]{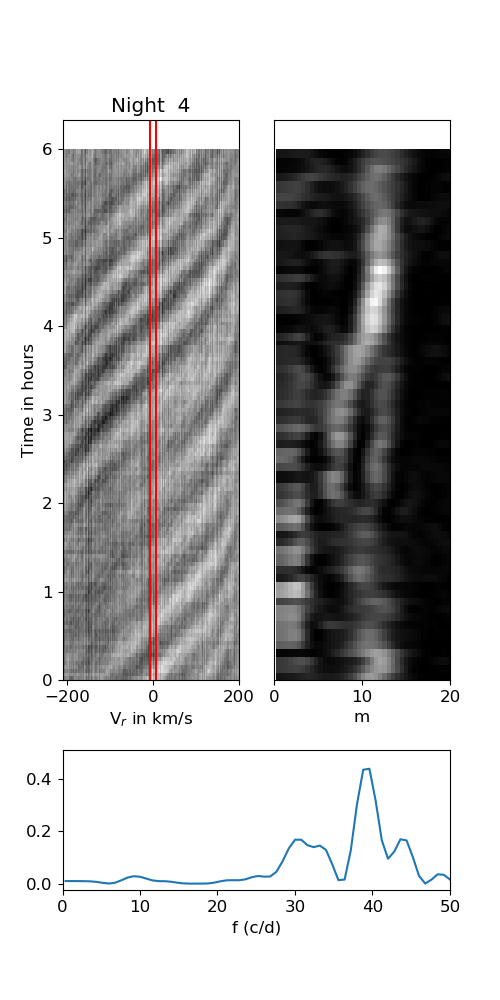}
    \includegraphics[width=0.30\linewidth]{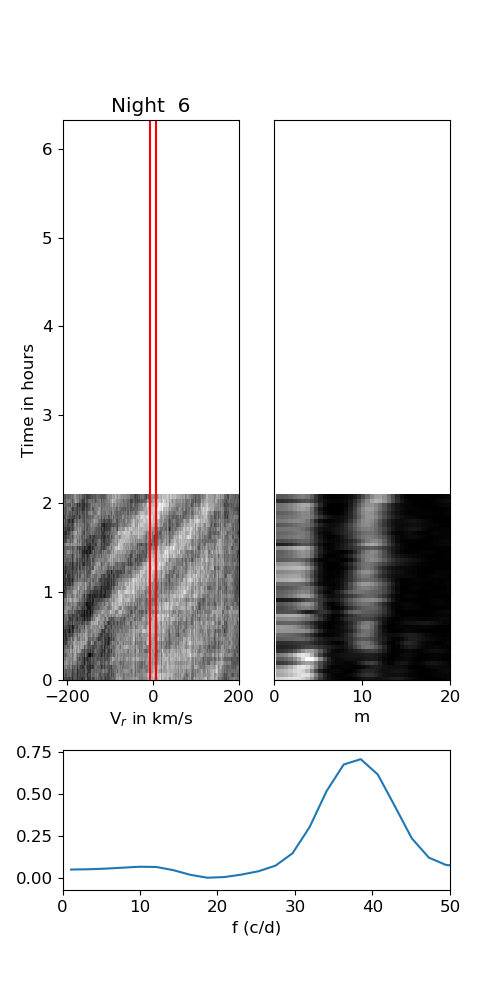}
    \includegraphics[width=0.30\linewidth]{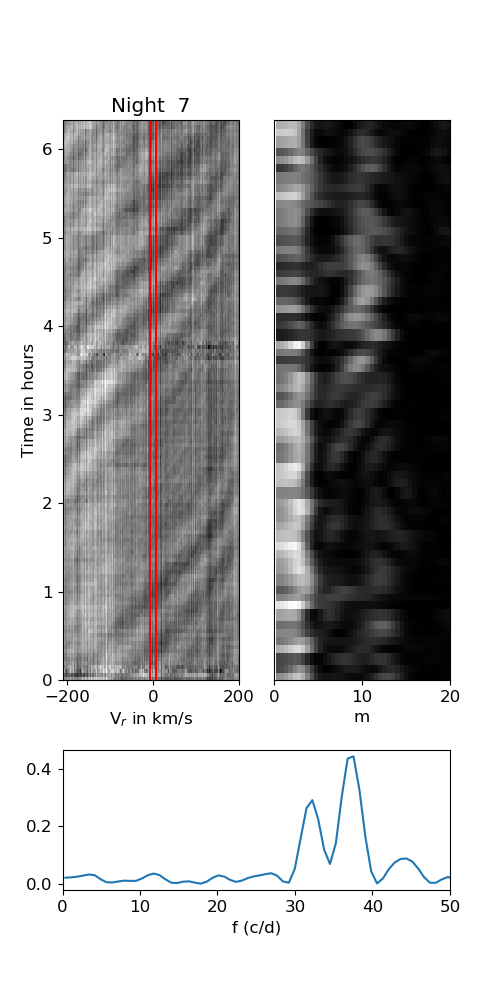}
    \caption{Global view of the relative fluctuations in the mean line profile
for each of the six nights of observation. For each night we show in
the top left panel the signal as a function of time and radial velocity
in km/s. In the top right panel we show the Lomb-Scargle periodogram
of the line profile fluctuations in (virtual) longitude as a function
of time and the azimuthal wavenumber $m$, while the bottom plot shows a
Lomb-Scargle periodogram of the time variations of the intensity in the
eight pixels at the line centre that are marked by the red lines in the
top-left panels.  }
    \label{globalview}
\end{figure*}

Relative amplitudes of the detected features are typically of order $10^{-3}$
in intensity (Fig.~\ref{mean_prof} bottom) and look like wave trains
propagating in the prograde direction as shown in Fig.~\ref{globalview}'s
left plots for all the nights. Typically, individual wave trains remain
visible for roughly two hours.


We shall assume that these waves only propagate in longitude and that
they are of the form $I(\theta)\exp(im\varphi-i\omega t)$, where
$(\theta,\varphi)$ are the spherical angles of a point at Altair's
surface.  If perturbations are concentrated around Altair's equator
we can easily relate the radial velocity $V_r$ at which the perturbation
occurs and the longitude by

\beq \varphi_v = \arcsin\lp\frac{V_r}{V_{\rm eq}\sin i}\rp\eeq
where $V_{\rm eq}$ is the equatorial velocity and $i$ is the angle
between the line of sight and the rotation axis. Of course $V_r$ is
corrected from the radial
velocity of Altair.
We shall call $\varphi_v$ the virtual longitude to stress the fact that
actually several longitudes contribute at a given radial velocity as is
clear from the projected map shown in Fig.~\ref{vproj}.

\begin{figure*}
    \centering
    \includegraphics[width=0.30\linewidth]{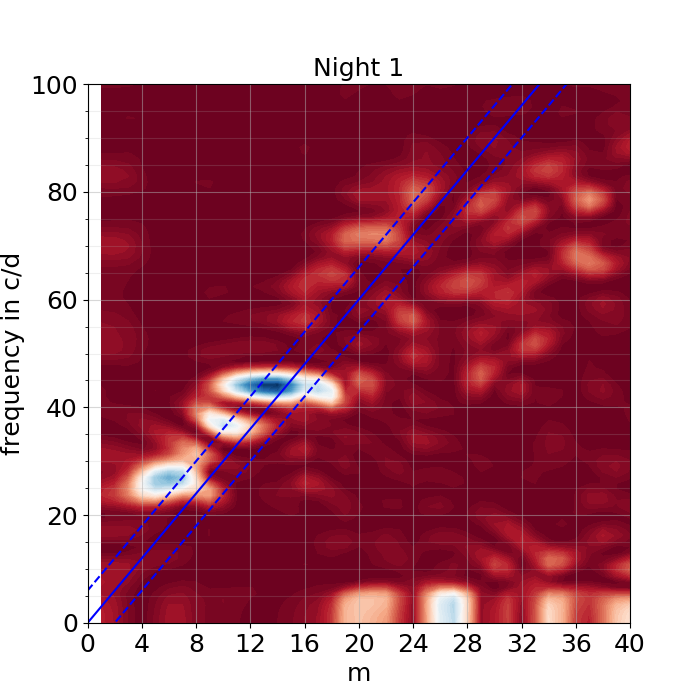}
    \includegraphics[width=0.30\linewidth]{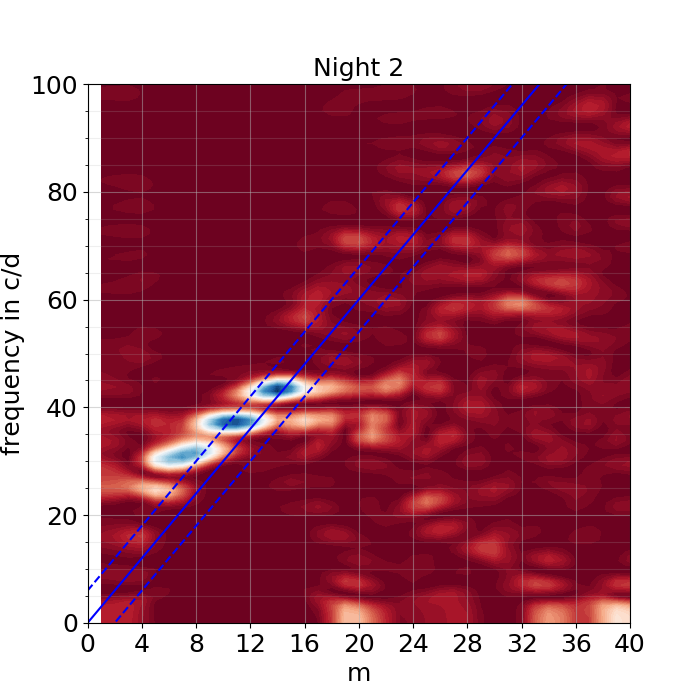}
    \includegraphics[width=0.30\linewidth]{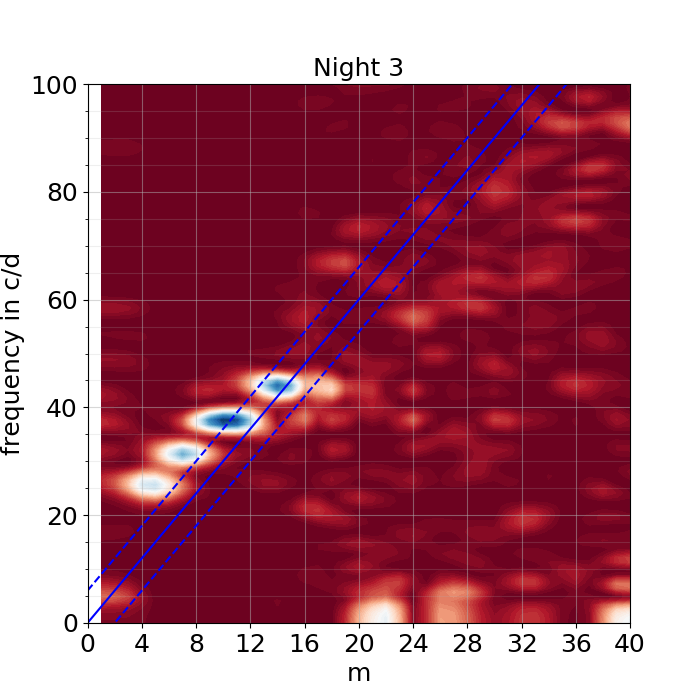}
    \includegraphics[width=0.30\linewidth]{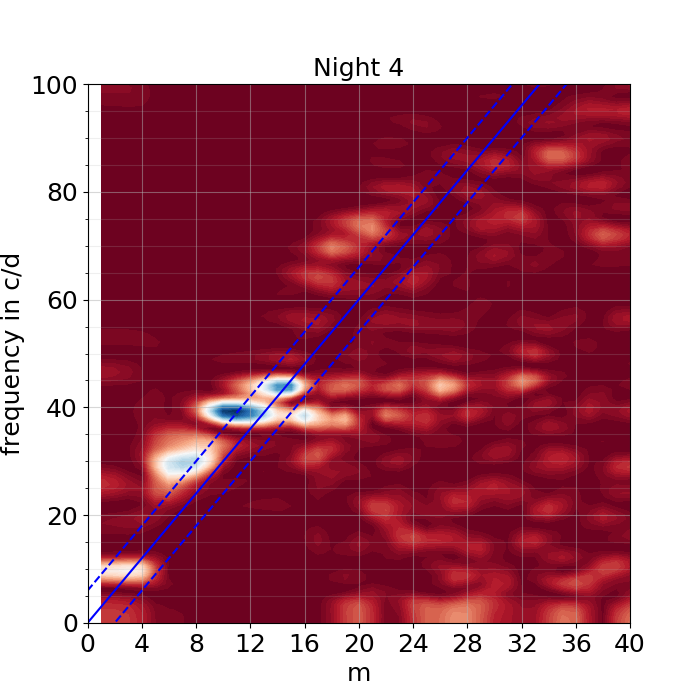}
    \includegraphics[width=0.30\linewidth]{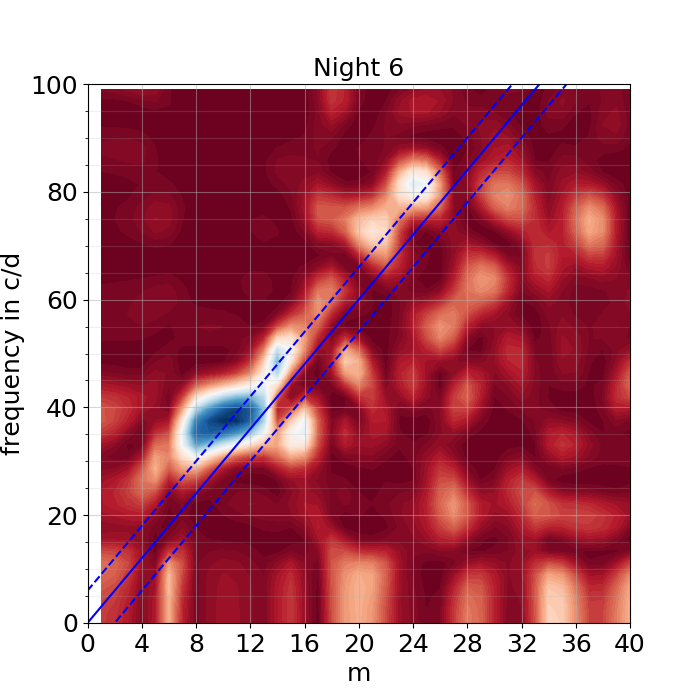}
    \includegraphics[width=0.30\linewidth]{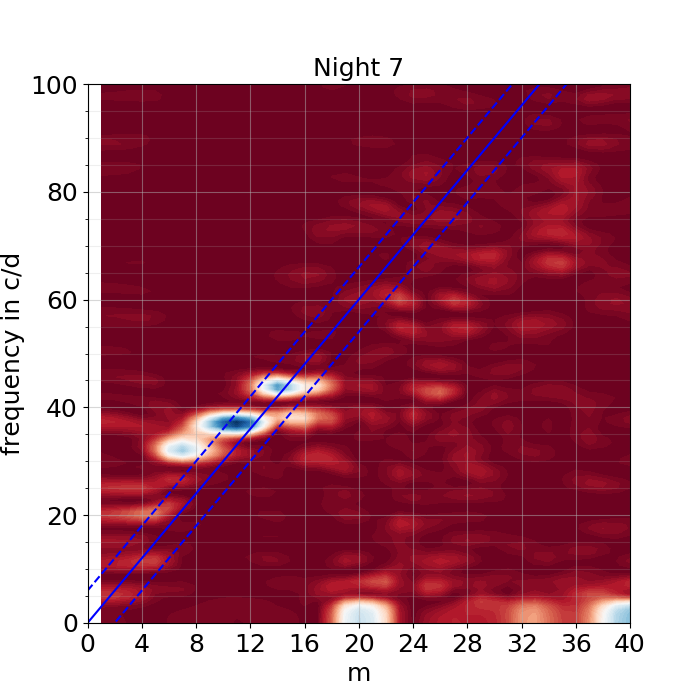}
    \caption{$m-\omega$ diagram for the six nights showing the spectral
power as a function of $m$ and frequency in the observer's
frame. The blue solid line shows the time frequency of a standing wave in the
corotating frame characterized by wavenumber $m$. Modes above that line
are prograde, and modes below that line are retrograde. The two dashed
blue lines delineate the inertial frequency band.}

    \label{m-omega}
\end{figure*}

It is nevertheless interesting to see the azimuthal wavenumber $m$ which
is associated with the perturbations of the line if we replace the radial
velocity coordinate by the virtual longitude. Hence, we get an
information on the apparent $m$ values contained in the signal
\cite[e.g.][]{bohm+09}.
To this end, we compute a Lomb-Scargle
periodogram of the line profile fluctuations (like the one shown in
Fig.~\ref{mean_prof} bottom) using the virtual longitude
as the time analog\footnote{Lomb-Scargle
periodogram are usually computed for time-dependent signals.}. We choose the Lomb-Scargle periodogram instead of
a simple Fourier transform since the data points are not regularly spaced
in longitude (see below). This computation allows us to extract the azimuthal
wavenumbers contained in the signal as a function of time as shown
in Fig.~\ref{globalview} (right plots). These plots clearly show that
wavenumbers around $m=10$ are conspicuously detected. We also note that
wavenumbers change with time. It may be a consequence of beating waves
or some nonlinear evolution.

Beside the foregoing azimuthal analysis of the line profile
fluctuations, we can also compute a Lomb-Scargle periodogram at a
given virtual longitude. We thus calculate this periodogram selecting
eight pixels around the line centre and thus exhibit for each
night a few frequencies. The corresponding plots are still shown in
Fig.~\ref{globalview} at the bottom for each night. There we clearly
see that all detected frequencies lie in between $\sim20$ 
and $\sim50$ cycles/day (c/d).

The next step is to relate the frequencies and the azimuthal wavenumbers
$m$. For that purpose we first fit each line profile fluctuation with a
limited Fourier series, namely

\beq \delta I(t,\varphi_v) = a(t)+
\sum_{m=1}^{20}b_m(t)\cos(m\varphi_v)+c_m(t)\sin(m\varphi_v) \eeq
We then compute the Lomb-Scargle time periodogram of each coefficient
$b_m(t)$ and $c_m(t)$, for all the chosen wavenumbers. We thus derive
the transforms $\tilde b(m,\omega)$ and $\tilde c(m,\omega)$, from
which we compute the spectral power $P(m,\omega) = \sqrt{\tilde
b^2+\tilde c^2}$. As a result we obtain an $m-\omega$ diagram
analogous to the famous $k-\omega$ diagram of solar eigenmodes
\cite[e.g.][]{gonzalez-hernandez+98,gizon+10}. This diagram, showing
$P(m,\omega)$, is displayed for each night in Fig.~\ref{m-omega}.

As may be noted, spatial frequencies range from $m=5$ to $m\simeq18$
while time-frequencies stay in between 20~c/d and 50~c/d.  The blue line
on each diagram shows the frequency of a standing wave of azimuthal
wavenumber $m$ in the corotating frame, assuming a rotation period
of 8 hrs deduced from the model of Altair by \cite{bouchaud+20}. We
note that most of the modes are above that line showing that they are
prograde modes in the corotating frame. Moreover, they are mostly in the
inertial frequency band
$-2f_{\rm rot}\leq f_{\rm corot} \leq2f_{\rm rot}$ or slightly above, thus
indicating their nature, namely inertial or gravito-inertial modes
\cite[e.g.][]{RV97,DRV99,DR00}.

\begin{table}[t]
    \centering
\begin{tabular}{|lll|c|c|}
 \multicolumn{3}{|c|}{Frequency} & $\moy{A}$ & $m$ \\
\multicolumn{3}{|c|}{(c/d)}   &    (ppm)   &      \\
  & &                   &  \\
$f_1$ &  $38.14$ & 4.26 & 294$\pm7$ & { 11} \\
$f_2$ &  $43.63$ & 3.59 &210$\pm6$ & { 13} \\
$f_3$ &  $31.51$ & 6.87 & 166$\pm7$ & { 8}  \\
$f_4$ &  $25.32$ & 6.84 & 123$\pm6$ & { 6}  \\
$f_5$ &  $35.87$ & 1.99 & 113$\pm7$ & { 11}\\
$f_6$ &  $31.13$ & 6.49 & 102$\pm6$ & { 8}\\
$f_7$ &  $56.58$ & 10.38&  27$\pm4$ & { 15}\\
   \multicolumn{5}{c}{}
\end{tabular}
\caption{List of oscillation frequencies that have been
detected with their most probable azimuthal wavenumber ($m$). The first
column of the frequencies are the one observed, while the second column
gives the value in the corotating frame of Altair using a rotation
frequency of 3.08 c/d \cite[e.g.][]{ledizes+21}. $\moy{A}$ is the measured
amplitude.}
    \label{freq_altair}
\end{table}

\begin{figure}
\includegraphics[width=0.95\linewidth]{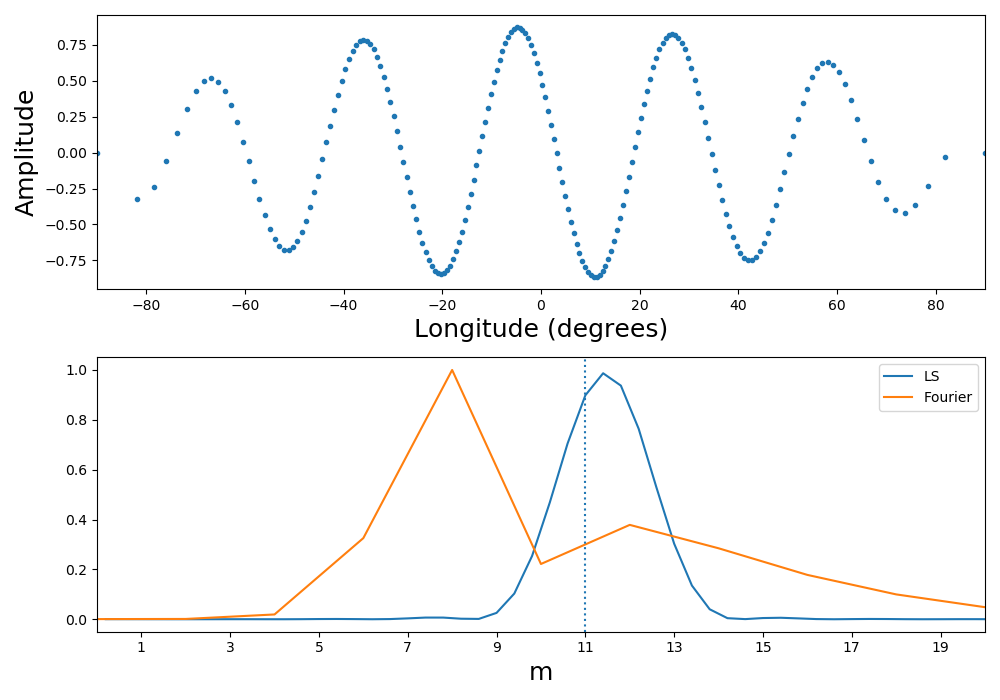}
\caption{Top: Simulation of the line profile perturbation induced by a
purely sectoral mode with $m=11$ (amplitude is arbitrary). Bottom: the Fourier
spectrum of the signal and the Lomb-Scargle periodogram.}
\label{simu}
\end{figure}

\begin{figure}
\includegraphics[width=0.95\linewidth]{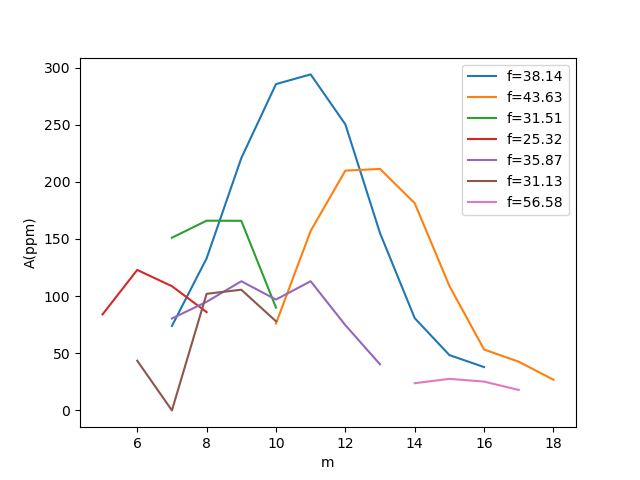}
\caption{Amplitude (in ppm) of the waves as a function of the azimuthal
wavenumber used to do the projection of the Doppler signal. Waves are
characterized by their frequency $f$ given in c/d.}
\label{ampl}
\end{figure}

\begin{figure*}
\centering
\includegraphics[width=0.49\linewidth]{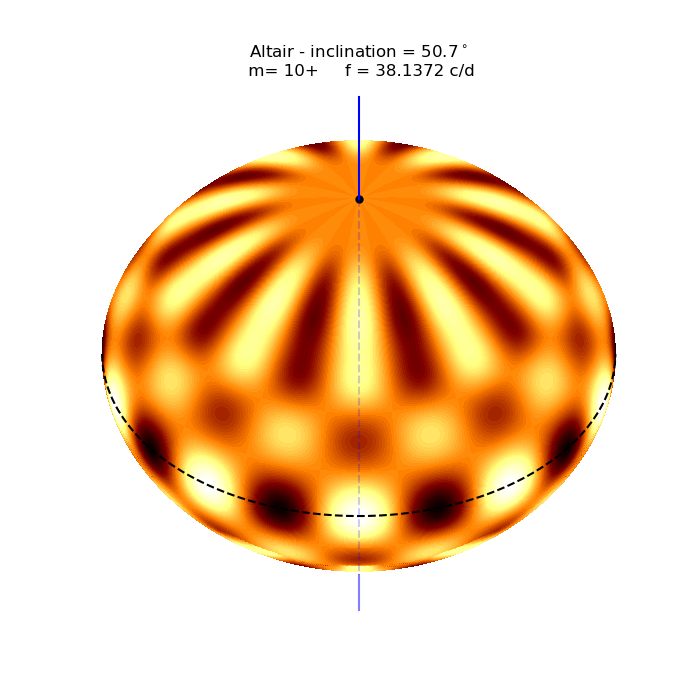}
\includegraphics[width=0.45\linewidth]{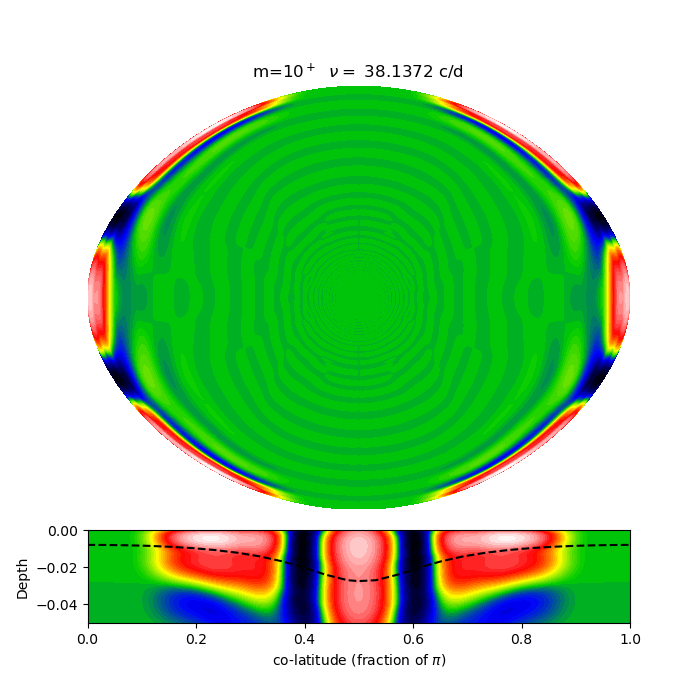}
\caption{Surface amplitude (left) and meridional cut of the kinetic
energy density (right) of a gravito-inertial mode that may give a signal
at f=38.14 c/d. The bottom plot at right is a zoom of the surface layers
and the dashed line shows the T=50,000~K isotherm around which the second
ionization of helium takes place. The depth is scaled by the polar radius
of the model.}
\label{surf3814}
\end{figure*}

\begin{figure}
\includegraphics[width=1.00\linewidth]{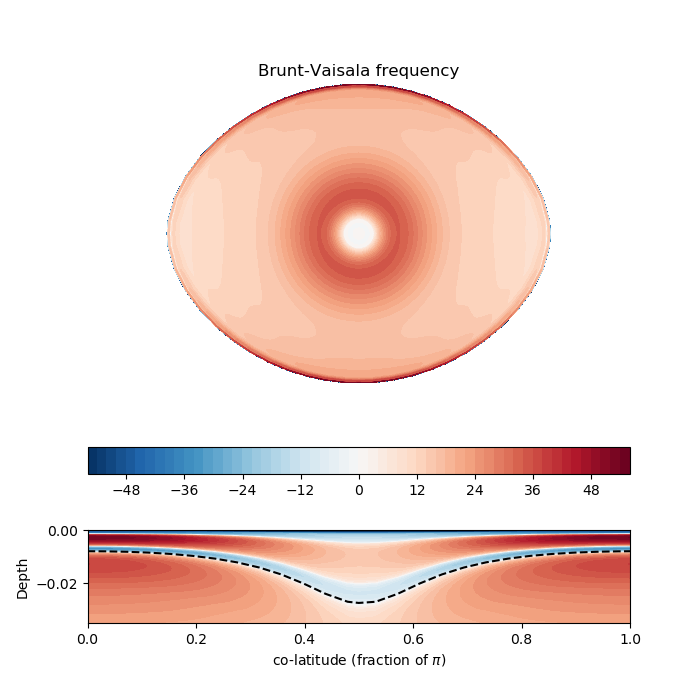}
\caption{Meridional section of the concordance model of Altair
\cite[e.g.][]{bouchaud+20} showing the \BVF. Units on the color bar are
in cycle/day. Negative values show the convectively unstable regions.
The lower plot gives a zoomed-view of the surface layers and the dashed line shows the T=50,000~K isotherm around which the second
ionization of helium takes place. Depth is
scaled by the polar radius of the model.}
\label{BV}
\end{figure}

\section{Mode detection and identification}

The foregoing raw analysis showed that a set of waves are excited. To
proceed towards their identification we first try to identify their
azimuthal wavenumbers of their longitude dependence, namely their
$\exp(im\varphi)$-dependence. As illustrated in Fig.~\ref{simu} (top) where
we show the spectral perturbation generated by a purely sectoral mode,
propagating over \cite{bouchaud+20}'s model, with a latitude-longitude
amplitude fluctuation like

\beq
\delta a \propto \sin^m\theta\exp(im\varphi),
\eeq
with $m=11$,
the signal is neither periodic nor evenly sampled in
longitude. Hence, the Fourier transform struggles to recover the
right $m$. Fig.~\ref{simu} shows that the Lomb-Scargle periodogram is
more appropriate than the Fourier spectrum to recover the actual $m$ of
the signal.

To further progress, we compute the Lomb-Scargle periodogram of $b_m(t)$
and $c_m(t)$ combining all the nights' data. As expected from the
$m-\omega$ diagram (Fig.~\ref{m-omega}), we detect the same frequency for
several $m$'s, but with different amplitudes. Fig.~\ref{ampl} summarizes
the results and shows the amplitude as a function of $m$ for each given
frequency. Obviously the most prominent wave oscillates at 38.14 c/d
and seems to be associated with a $m=11$ wavenumber.  In
Tab.~\ref{freq_altair}, we list the frequencies that have been
unambiguously detected. We also give the amplitude of the signal in ppm
and the most probable azimuthal wavenumber $m$. As shown by
Fig.~\ref{ampl}, the association between frequency and wavenumber is not
always clear.

Three of the detected frequencies $(f_1,f_2,f_5)$ are clearly in
the inertial frequency band, since they verify \mbox{$f_{\rm corot}\leq
2f_{\rm rot}$}, if we take $f_{\rm rot}=3.08$~c/d (see Sect. 3). The
four other frequencies are also low frequencies only slightly above
the inertial band. The modes associated with these frequencies are
therefore gravito-inertial modes which may be either inertial modes
(restored by Coriolis force) perturbed by a stable stratification
or reciprocally, gravity modes perturbed by rotation \cite[see][for
canonical examples]{DRV99}.

\section{The word of models}

\subsection{Preliminary results}

To have further insight into the waves that are seen in
the present spectroscopic data, we now focus on the concordance model of
Altair derived by \cite{bouchaud+20}, which we previously mentioned. We
recall in Tab.~\ref{estermodel} the fundamental parameters of this
model that matches the constraints derived from interferometric,
spectroscopic and seismic data. The latter data are frequencies
obtained from WIRE photometry by \cite{buzasietal05}.

\begin{table}
\centering
\begin{tabular}{|c|c|}
 & \\
Parameters & ESTER model \\
 & \\
$M$ (\msun)                & 1.863 \\
$\Tpol$ (K)                & 8621\\
$\Teq$ (K)                 & 6780\\
$\Rpol$ (\rsun)            & 1.568 \\
$\Req$ (\rsun)             & 2.011 \\
$\veq$ (km/s)              & 313\\
$\Omega_{\rm eq}$ ($\Omega_k$)      & 0.744 \\
$\varepsilon$              & 0.220 \\
$Z$                        & {0.0192}\\
$X_{\rm env}$              & {0.739}\\
$X_{\rm core}$             & {0.712}\\
 & \\
\end{tabular}
\caption{Fundamental parameters of an ESTER model that matches the
observational constraints on Altair derived by \cite{bouchaud+20}.
$\varepsilon=1-\Rpol/\Req$ is the flattening, $Z$ is metallicity and $X$
the hydrogen mass fraction.}
\label{estermodel}
\end{table}

To further progress in the seismological properties of Altair, we
computed some eigenmodes of the concordance model with the TOP code 
\cite[][]{reese+21}, which can handle ESTER 2D models. We naturally focus
on the observed frequencies but more precisely, we scanned the frequency
band $f\pm0.05$~c/d of each frequency listed in Tab.~\ref{freq_altair}
for the three $m$'s around the most probable one. Hence, for the most
prominent frequency, at 38.14 c/d, we investigate the frequency band
$[38.09, 38.19]$ c/d for azimuthal wavenumbers $m=10, 11, 12$. We show in
Fig.~\ref{surf3814} a view of a mode which may give the signal
observed at 38.14 c/d.
Shifting this frequency in the co-rotating frame, namely

\beq f_{\rm corot} = f - m f_{\rm rot}\; ,\eeq
with $f_{\rm rot}=3.08$~c/d, we find that it either belongs
to the inertial frequency band, if $m=11$ or 12, or is just above if
$m=10$. As shown in Fig.~\ref{surf3814}, the amplitude of such a  mode
is only significant near the surface. The meridional map of the \BVF\
(Fig.~\ref{BV}) shows that such modes actually propagate over two
convectively unstable layers sandwiching a stable one. The deeper
unstable layer is associated with the second ionization of helium,
which is the driver of the kappa-mechanism in $\delta$ Scuti stars
\cite[][]{baglin+73,balona+15}. The 50,000K isotherm, around which
the second ionization of helium takes place, is depicted as a dashed line in
Fig.~\ref{surf3814} (bottom right) and Fig.~\ref{BV} (bottom), showing
that this mode may possibly be destabilized by the kappa mechanism.

The foregoing mode may be used to constrain the differential rotation
of the star in the 1\%-depth surface layers.  The knowledge of this
differential rotation, close to the surface, will help the modelling of
a dynamo, which may be at the origin of the X-ray activity of
Altair \cite[][]{robrade+09}.

\subsection{Discussion}

The mode shown in Fig.~\ref{surf3814} has been selected because of its
(presumably) high visibility but we still ignore whether it is stable or
not. Indeed, the foregoing computation neglected any non-adiabatic
effect since preliminary non-adiabatic calculations showed inconclusive
results for many reasons that we shall discuss now.

Indeed, gravito-inertial modes
form a dense spectrum in the adiabatic limit. Namely, any frequency
below $2f_{\rm rot}$ is as close as we wish to a mode frequency
\cite[e.g.][]{DRV99}. Moreover, the modes contain singularities, which appear
as shear layers in the eigenfunctions \cite[e.g.][]{DRV99,RGV01,RV18}. This
makes the computation of gravito-inertial modes in a stellar model quite
difficult. In particular, our attempts could not reliably
compute the growth or the damping rates of the eigenmodes with
frequencies in the range of the observed ones (e.g.
Table~\ref{freq_altair}).

If we leave aside the stability question, the density of frequencies in
the spectrum is another difficulty. The observed frequencies
listed in Tab.~\ref{freq_altair}
have a limited precision of 0.05 c/d due to the short length of the
time series. Our calculations show that in a frequency box of size
0.1~c/d a dozen of eigenmodes may be found if we only consider the
least damped modes. Obviously, long time series will be needed to reduce
this uncertainty.

On the theoretical side, the instability that drives the modes may not
be of the usual nature, namely an exponential growth that is limited
by a nonlinear coupling with stable modes. Here, the spectral density
of modes reveals the non-normality of the differential operator which
governs the free oscillations.
It allows the possibility of algebraic growth of perturbations
\cite[][]{schmid07,rieutord15}. In such a case the observed waves are
not eigenmodes, but combinations of several eigenmodes.

The foregoing remarks underline the point that deciphering the waves
frequencies at the surface of Altair will be a challenging task to which
we reserve a dedicated work.

On the observational side, it is clear that long time series are needed
to narrow the error box on frequencies and thus be of great help to
identify the possible modes or quasi-modes that are observed.

\section{Conclusions}

In this paper we presented the first detection by spectroscopy of waves
at the surface of Altair, a rapidly rotating star with an equatorial
velocity likely over 300~km/s. We showed that the observed waves are in or near
the frequency band $[-2f_{\rm rot},2f_{\rm rot}]$ in the frame co-rotating with
the star. They are thus identified as inertial or gravito-inertial waves.
The difference between inertial waves and gravito-inertial ones comes from
the influence of the stable stratification of the fluid. Pure inertial
waves are restored only by the Coriolis force and, in stars, appear in
their convection zone \cite[e.g. the recent detection of these waves in
the Sun by][]{gizon+21}, while  gravito-inertial waves are restored both
by buoyancy and Coriolis force. In Altair, thin convective layers exist
close to the surface and may be the seat of pure inertial waves, which
may drive a signature at the surface. An identification of observed waves
with gravito-inertial waves is however more likely since convective
layers are thin and separated by a radiative one.  Finally, we note that the
observed waves are characterized by rather high azimuthal wavenumbers,
and propagate in the prograde direction. Their amplitude, in intensity,
is of the order of 10$^{-3}$.

Altair was known as the ``brightest $\delta$ Scuti"
\cite[][]{buzasietal05}. We now see that its oscillation spectrum not only
includes acoustic modes but also gravito-inertial waves. Low-frequency
oscillations were actually already detected by \cite{buzasietal05}
and \cite{ledizes+21} in photometric data, but without constraints on
the wavenumbers it was difficult to assign them a definite category
(gravito-inertial, pure inertial or even acoustic) because of the high
rotation frequency ($\sim 3$c/d) and a possible non-axisymmetric nature.
Hence, Altair now appears as a hybrid oscillator but its fast rotation
makes it still different from stars that show the hybrid state of
$\delta$ Scuti and $\gamma$ Doradus stars.

Our failure to identify more precisely the observed frequencies,
using the best model of Altair \cite[][]{bouchaud+20}, shows that such
an identification is difficult due to the spectral density of modes in
or near the inertial frequency range. A dedicated work is therefore required to
further progress in the interpretation of the observed frequencies, all
the more that gravito-inertial modes  are controlled by a non-normal
differential operator in the adiabatic limit. This implies that the growth
of the mode may be algebraic like in shear instabilities, and may excite
a wide or ever changing oscillation spectrum \cite[][]{schmid07}.

On the observational side, further progress claims for longer time series
to either give a more precise value to the frequencies, and to eventually
monitor the presently detected oscillations. Moreover, it will be very
interesting to observe spectroscopically other fast rotators similar to
Altair to see if such waves are also excited in stars with different
fundamental parameters.  $\alpha$ Ophiuchi (Ras Alhague) and $\alpha$
Cephei (Alderamin) are very good targets for such investigations because
of their brightness, but the recent work of \cite{ma+22} shows that
Altair may have numerous sisters.

\bigskip
\begin{acknowledgements}
MR would like to thank S. Charpinet for enlighting discussions on the
analysis of the time series. He also acknowledges the support of the French
Agence Nationale de la Recherche (ANR), under grant ESRR
(ANR-16-CE31-0007-01). GMM ackowledges support by ``Contribution of
the UGR to the PLATO2.0 space mission. Phases C / D-1",1032 funded
by MCNI/AEI/PID2019-107061GB-C64. Computations of Altair's models and
eigenfrequencies have been possible thanks to HPC resources from CALMIP
supercomputing center (Grant 2022-P0107).

\end{acknowledgements}

\bibliographystyle{aa}
\bibliography{../../../biblio/bibnew}
\end{document}